\documentclass[aps,prd,reprint,onecolumn,notitlepage,superscriptaddress]{revtex4-1}

\usepackage{lipsum}
\usepackage{amsmath}
\usepackage{graphicx}
\usepackage{color}
\usepackage{tikz}
\usepackage{verbatim}
\usepackage{hyperref}
\usepackage{amssymb}
\usepackage{bm}
\graphicspath{ {./}}

\newcommand{\rom}[1]{\textup{\uppercase\expandafter{\romannumeral#1}}}

\begin{document}

\title{Self-healing unitarity is an optical illusion}

\author{Archit Vidyarthi \footnote{email:archit17@iiserb.ac.in}
       }

\affiliation{Department of Physics, Indian Institute of Science Education and Research Bhopal, Madhya Pradesh - 462066, India}

\begin{abstract}
Among the vast variety of proposals put forward by the community to resolve tree-level unitarity violations in Higgs inflation models, there exists the concept of self-healing. It heals the theory from supposed tree-level violations for elastic scattering processes by summing over successive vacuum polarization loop corrections. In this work, we examine this technique to check whether unitarity is indeed restored and find that there exist underlying constraints in self-healing unitarity that pose the same perturbative unitarity bounds that it was expected to heal.

\end{abstract}

\maketitle

\section{Introduction}\label{intro}
Unitarity is one of several properties at the heart of a quantum theory, and essentially implies that the probability of an event cannot exceed unity. Along with other properties such as positivity, causality, etc., it helps provide us with useful bounds on a theory (for example: perturbative bounds, Froissart bounds, etc. \cite{Schwartz:2014sze}) in the form of constraints on a parameter, or on the domain within which the theory is valid, without needing to introduce new degrees of freedom (DsOF).

Tree-level unitarity violations, estimated using perturbative unitarity bounds, are immensely helpful in pointing out missing pieces in a theory. For a non-renormalizable theory, these may imply that the loop corrections might become relevant as we approach the apparent violation scale in describing the complete process \cite{Schwartz:2014sze}. For others, they may indicate that the theory is incomplete. Beyond Standard Model (BSM) physics helps fill in gaps stemming from the incompatibility of the Standard Model and gravity, and provides us with possible candidates for the missing DsOF, often motivated by the existence of dark matter and dark energy that make up the majority of the energy content of the universe. 

Given how Higgs driven inflation has been one of the prime candidates for an inlfaton field (check \cite{Atkins:2010yg,Lerner:2009na,Rubio:2018ogq} and references therein), the fact that it faces unitarity violations far below the Planck scale is something the scientific community has been trying to explain away for a long time (see \cite{Panda:2022esd,Calmet:2013hia,Rubio:2018ogq,Lerner:2010mq,Escriva:2016cwl,Antoniadis:2021axu} and references therein for more info). After several decades of search, though, we have as of yet not been able to resolve the issue completely. Among the several approaches suggested towards resolving the issue is self-healing of unitarity proposed in \cite{Aydemir:2012nz} and later applied in the context of Higgs inflation in \cite{Calmet:2013hia}, which are at the heart of what we discuss in this work.

This paper is organized as follows: in Sec.\ref{recap}, we introduce the reader to the optical theorem and partial wave unitarity bounds as presented in \cite{Schwartz:2014sze}; in Sec.\ref{prop}, we briefly review the idea of self-healing as it was put forward in \cite{Aydemir:2012nz} while briefly introducing \cite{Han:2004wt,Calmet:2013hia}; in Sec. \ref{viola} we critically examine \cite{Aydemir:2012nz,Calmet:2013hia} and assess whether self-healing unitarity mechanism does play a role in the context of unitarizing nonminimally coupled scalar-tensor theories (STTs); and lastly, we conclude in Sec.\ref{concl}.

\section{Perturbative Unitarity Bounds}\label{recap}
\begin{figure}
   \centering
    \includegraphics[width=6.5cm, height=4cm]{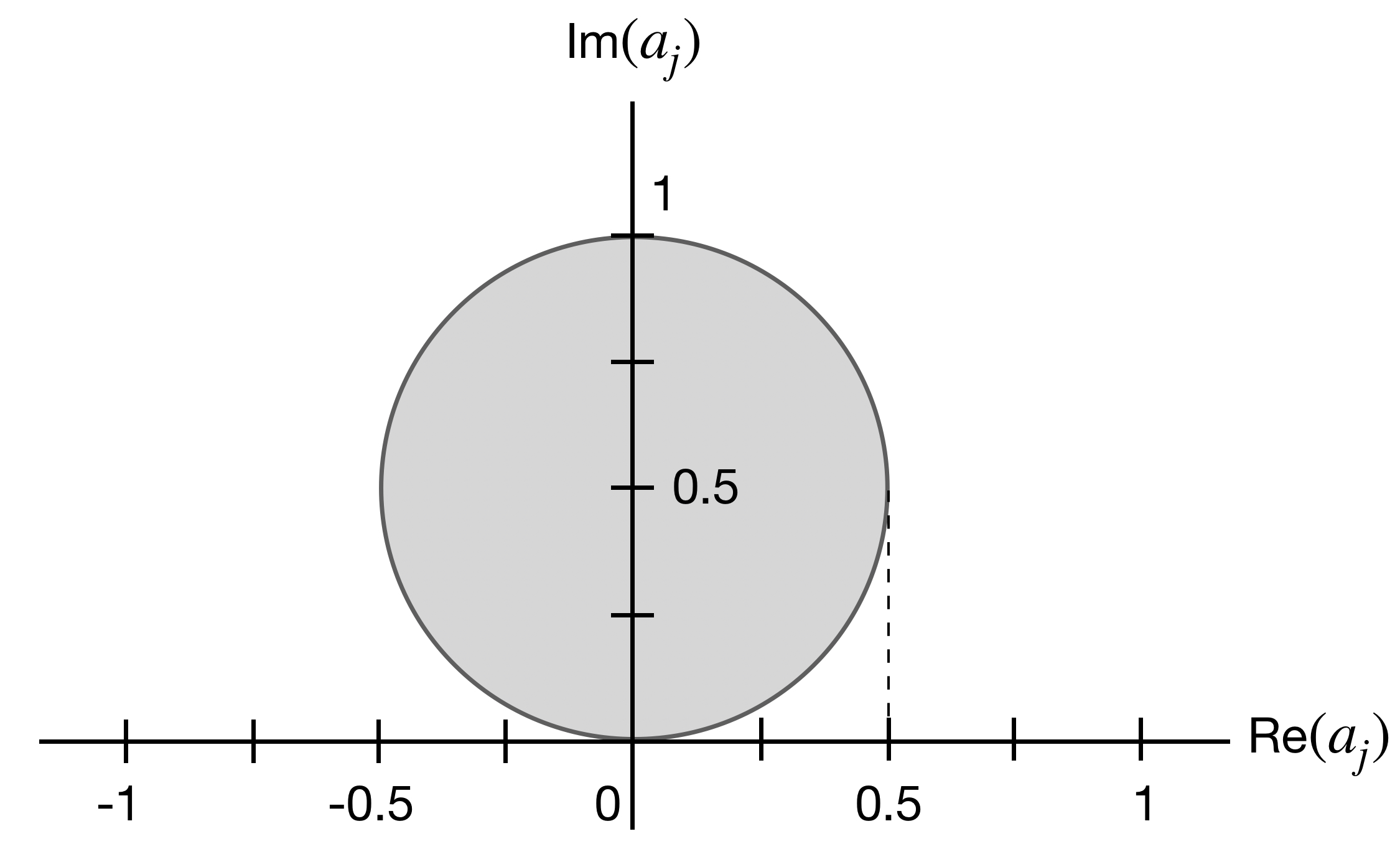}
    \caption{Argand diagram representing the condition $|a_j|^2\leq \text{Im}(a_j)$. The area within the circumference represents the acceptable region where unitarity is obeyed.}
    \label{fig:1}
\end{figure}

Imposing that the action is unitary, we obtain the famous optical theorem, which equates the imaginary part of the scattering amplitude to the total scattering cross section. 
\begin{equation}\label{opt}
    \mathcal{M}(i\to f)-\mathcal{M}^*(f\to i)=i\sum_X\int d\Pi_X (2\pi)^4\delta^4(p_i-p_X)\mathcal{M}(i\to X)\mathcal{M}^*(f\to X).
\end{equation}
where $\mathcal{M}$ represents the scattering amplitude, $\left|i\right>$, $\left|f\right>$, $\left|X\right>$ are initial, final and arbitrary intermediate states, respectively, $p_n$ represents the momentum of state $\left|n\right>$, and $d\Pi_X$ is the momentum integral measure. In its generalized form (\ref{opt}), this theorem states that order-by-order in perturbation theory, imaginary parts of higher loop amplitudes are determined by lower loop amplitudes. For instance, the imaginary part of one-loop amplitude could be determined by the tree-level amplitude. A special case arises from this using the assumption that the initial and final states are the same (i.e. $\left|i\right>=\left|f\right>=\left|A\right>$):
\begin{equation}\label{init}
    \text{Im}\mathcal{M}(A\to A)=2E_{CM}|\Vec{p}_i|\sum_X\sigma(A\to X).
\end{equation}
where $E_{CM}$ is the center of mass energy of the system and $\sigma(A\to X)$ is the scattering cross section for the enclosed process. Optical theorem puts a constraint on how large a scattering amplitude can be. From the approximate form,
\begin{equation}
    \text{Im}\mathcal{M}\leq|\mathcal{M}|^2\implies|\mathcal{M}|<1.
\end{equation}
Now, using the partial wave expansion of the scattering amplitude to impose constraints on coefficients of the Legendre polynomials. To recap, we first expand the scattering amplitude as:
\begin{equation}
    \mathcal{M}(\theta)=16\pi\sum_j a_j (2j+1) P_j(\cos{\theta}),
\end{equation}
where $a_j$ are complex-valued coefficients, and $P_j(\cos{\theta})$ are Legendre polynomials with $P_j(1)=1$ and
\begin{equation}
    \int_{-1}^{1} P_j(\cos{\theta}) P_k(\cos{\theta}) d\cos{\theta}=\frac{2}{2j+1}\delta_{jk}.
\end{equation}
For a case where the initial and final states are the same, we can write the total scattering cross section in the center of mass frame as:
\begin{equation}
    \sigma_{CM_{tot}}= \frac{16\pi}{E_{CM}^2}\sum_j |a_j|^2 (2j+1).
\end{equation}
Employing the optical theorem at $\theta=0$, we have,

\begin{equation}\label{ineq}
\text{Im} \mathcal{M}(A B \rightarrow A B \text { at } \theta=0) =2 E_{\mathrm{CM}}\left|\vec{p}_i\right| \sum_X \sigma_{\mathrm{tot}}(A B \rightarrow X) \geq 2 E_{\mathrm{CM}}\left|\vec{p}_i\right| \sigma_{\mathrm{tot}}(A B \rightarrow A B),
\end{equation}
where an inequality has been introduced owing to the fact that $\left|AB\right>\in \left|X\right>$. Then,
\begin{equation}\label{partial}
\sum_{j=0}^{\infty}(2 j+1) \operatorname{Im}\left(a_j\right) \geq \frac{2\left|\vec{p}_i\right|}{E_{\mathrm{CM}}} \sum_{j=0}^{\infty}(2 j+1)\left|a_j\right|^2 .
\end{equation}
This, coupled with the inequality $|a_j|\geq \text{Im}(a_j)$, means that the magnitude of $a_j$ is now constrained as $|a_j|\leq1$, $0\leq\text{Im}(a_j)\leq 1$, and $|\text{Re}(a_j)| \leq 1/2$. These three conditions constitute the perturbative unitarity bounds and can be found from the Argand plane in Fig.(\ref{fig:1}). The information presented in this section is explained in much greater detail in \cite{Schwartz:2014sze}.

\section{Self-healing Unitarity}\label{prop}

\begin{equation}\label{taomod}
    S=\int d^4 x \sqrt{-g}\left[\frac{1}{16\pi G_N}R+\partial^\mu\Phi^\dagger\partial_\mu\Phi-m^2|\Phi|^2+\lambda R |\Phi|^2\right]
\end{equation}
Preceding \cite{Aydemir:2012nz}, authors of \cite{Han:2004wt} worked with a set of complex scalar fields nonminimally coupled with gravity as in Eq.(\ref{taomod}), and tried to estimate the scattering amplitude for the process $s\Bar{s}\to s'\Bar{s'}$, where they set $s\neq s'$ to make sure that only the $s$-channel graviton exchange diagram contributed to the process, and they could avoid collinear divergences in the $t$ and $u$ channels. They claimed that in the limit where the number of particles is large, the leading order loop corrections are successive vacuum polarization diagrams and that any tree-level violations could be fixed by considering such higher-loop corrections.

Using partial wave expansion as in the previous section, they estimated the scale of unitarity violations for $j=2$:
\begin{equation}\label{bound}
    s=\frac{20}{G_N N},
\end{equation}
where $s$ is a Mandelstam variable, $G_N$ is the Newton's constant and $N$ is the number of particles in the theory. They obtained the corresponding scales for the standard model of particle physics ($\sqrt{s}=6 \times 10^{18}GeV$) and the minimal supersymmetric standard model ($\sqrt{s}=4.6 \times 10^{18}GeV$), both coupled with gravity. 

Following this, authors of \cite{Aydemir:2012nz} considered a similar Lagrangian as \cite{Han:2004wt} involving a nonminimal coupling between gravity and multiple scalar fields and provide a useful confirmation for the results presented in \cite{Han:2004wt}. They go a step further and claim that the perturbative unitarity bound (\ref{bound}) is false, and that summing over the infinite loop contributions in the same manner as done in \cite{Han:2004wt}, they could use the relation:
\begin{equation}\label{infloopcond}
    a_j=\frac{a_j^{tree}}{1-\frac{Re(a_j^{1-loop})}{a_j^{tree}}-ia_j^{tree}}
\end{equation}
to show that there are no unitarity violations. Authors of \cite{Calmet:2013hia} expanded on this work and verified the results for a theory involving the Higgs' doublet They expanded the Higgs' boson around a large background which caused the coupling constants of the Higgs' and Goldstone bosons to differ. They, then, proceeded to show that self-healing phenomenon could be applied to $j=0$ level as well.


There are two equivalent approaches to observe self-healing of unitarity and we shall examine them both using results from \cite{Aydemir:2012nz,Calmet:2013hia} in the following section. It should be noted that even though the primary analyses have been performed in the Jordan frame, an equivalent self-healing mechanism can be found in the Einstein frame as well \cite{Ema:2019fdd}.

\section{Unitarity Violations in non-minimally coupled Scalar-tensor theories}\label{viola}
For this section, we consider the action:
\begin{equation}\label{stt}
    S=\int d^4 x \sqrt{-g}\left[\frac{1}{16\pi G_N}R+\partial^\mu\phi\partial_\mu\phi+\lambda R \phi^2\right]
\end{equation}
as used in \cite{Aydemir:2012nz} where we consider $N_s$ scalar fields and work in the large $N_s$ limit. Considering tree-level amplitudes, for $\lambda=0$ (minimal coupling) and $\lambda=-1/3$ (conformal coupling), the theory is well behaved up to the Planck scale, but more generally, looking at this action from dimensional grounds, we expect the perturbative unitarity scales for all processes $\phi_A\phi_A\to\phi_B\phi_B$ to be $\approx(G_N\lambda)^{-1}$. Now, for $\lambda>1$, it implies that perturbative unitarity is being violated below $\approx(G_N)^{-1}$. Authors in \cite{Han:2004wt} found that the perturbative unitarity bound in these theories was dependent on $N_s$, as stated in (\ref{bound}). The works \cite{Aydemir:2012nz,Calmet:2013hia} suggest, however, that the self-healing mechanism takes care of any supposed violations through infinite summation of vacuum polarization corrections.

\subsection{Partial wave amplitude approach}
The tree-level $\phi_A\phi_A\to\phi_B\phi_B$ scattering amplitude is,
\begin{equation}
    \mathcal{M}^{tree}=\frac{8\pi G_N}{s}[2 s^2\lambda(3\lambda+1)+ut]
\end{equation}
and corresponding partial wave amplitudes for all possible combinations of scalars can then be easily found to be,
\begin{align}
    a_0^{tree}=&\frac{G_N N_s s}{48}[24\lambda(3\lambda+1)+1],\label{treens}\\
    a_2^{tree}=&-\frac{G_N N_s s}{120}\label{treens1}.
\end{align}
Similarly, at 1-loop level,
\begin{align}
    \mathcal{M}^{1-loop}=&\frac{G_N^2 N_s}{15}[s^2F(\lambda)-ut]\\
    a_0^{1-loop}=&\frac{-G_N^2 N_s^2 s^2}{2880\pi}[6F(\lambda)-1]\ln{(-s)},\label{1lns}\\
    a_2^{1-loop}=&-\frac{-G_N^2 N_s^2 s^2}{14400\pi}\ln{(-s)}\label{1lns1}.
\end{align}
where $F(\lambda)=1+20\lambda+180\lambda^2+720\lambda^3+1080\lambda^4$ and $s$, $t$, $u$ represent the Mandelstam variables. The appearance of $N_s$ in the expressions (\ref{treens}), (\ref{treens1}), (\ref{1lns}), and (\ref{1lns1}) can be explained simply: the authors in \cite{Han:2004wt} worked with normalized two-particle states, such that the normalization factor for each state was $1/\sqrt{N_s}$. Combined with the combinatorial factor $N_s^2/2$ for large $N_s$, we are left with $N_s/2$ as the factor. For 1-loop results, we need to attach another $N_s$ factor for the scalars running in the loop. Note that $\lambda=0$ returns the expressions to the minimal coupling case, and $\lambda=-\frac{1}{3}$ gives us the conformal coupling case. For both of these, unitarity is safe up to the Planck scale. Since $a_2$ are completely independent of $\lambda$, the interesting case is clearly $a_0$. 

The authors in \cite{Aydemir:2012nz} only tackled the $j=2$ case and proved through $|a_2^{tree}|^2=\text{Im}(a_2^{1-loop})$ that the unitarity violations at tree-level can be taken care of by 1-loop corrections. It is clear from the expressions above, however, that since $j=2$ doesn't contain any $\lambda$ dependence, their results only prove that the minimal theory which appeared to have tree-level unitarity violations at the Planck scale is completely healed when considering the 1-loop corrections. This result was improved in \cite{Calmet:2013hia}, where the authors extended the result to the more relevant $j=0$ case. Their result holds even when we consider the coupling constants to be the same, as in (\ref{stt}).

The primary claim of \cite{Aydemir:2012nz} is that owing to the condition $|a_2^{tree}|^2=\text{Im}(a_2^{1-loop})$, we can write:
\begin{equation}\label{aycond}
    a_j=\frac{a_j^{tree}}{1+\frac{\text{Re}(a_j^{1-loop})}{a_j^{tree}}-ia_j^{tree}}.
\end{equation}
The authors state the same for (\ref{stt}) as well, which is corroborated and extended to $j=0$ in \cite{Calmet:2013hia}. They imply that perturbative unitarity violations have been dealt with completely and the dependence of violation scale on $N_s$ in \cite{Han:2004wt} is just a misinterpretation of the result. Now, as stated in Sec. \ref{recap}, the consequences of $|a_2^{tree}|^2=\text{Im}(a_2^{1-loop})$ are the constraints: $|a_j|\leq1$, $0\leq\text{Im}(a_j)\leq 1$, and $|\text{Re}(a_j)| \leq 1/2$. Considering the first constraint at tree-level (see appendix) for the partial wave amplitudes, we find,
\begin{align}
    |a_2^{tree}|\leq1&\implies s\leq \frac{120}{G_N N_s},\label{unlambound}\\
    |a_0^{tree}|\leq1&\implies s\leq \frac{48}{G_N N_s[24\lambda(3\lambda+1)+1]}\label{lambound},
\end{align}
i.e. there still exists a perturbative unitarity bound on the theory that depends on $N_s$ and $\lambda$, similar to the result of \cite{Han:2004wt} (up to a multiplicative factor). Therefore, we claim that the condition (\ref{infloopcond}) as proposed in \cite{Aydemir:2012nz} doesn't imply healed tree-level perturbative unitarity violations considering all loop levels, even though they appear to be healed when we consider 1-loop corrections. This can also be seen when we perform an summation over infinite vacuum polarization corrections. In order for the geometric progression to be convergent, we require that the common ratio be $<1$. This translates in our present case to the condition:
\begin{align}
    \left|\frac{a_2^{1-loop}}{a_2^{tree}}\right|&=\frac{G_N N_s s}{120\pi}<1\implies s<\frac{120\pi}{G_N N_s}\label{lambound1}\\
    \left|\frac{a_0^{1-loop}}{a_0^{tree}}\right|&=\frac{G_N N_s s}{60\pi}\left[\frac{6F(\lambda)-1}{24\lambda(3\lambda+1)+1}\right]<1\implies s<\frac{60\pi}{G_N N_s}\left[\frac{24\lambda(3\lambda+1)+1}{6F(\lambda)-1}\right]\label{unlambound1}
\end{align}
where we have ignored $\ln{(-s)}$ contributions since we're working in the UV limit. We see a similar dependence of the perturbative unitarity bound on $N_s$ and $\lambda$ as seen earlier in (\ref{unlambound}) and (\ref{lambound}).

\subsection{Dressed propagator approach}
Further, the authors in \cite{Aydemir:2012nz} verify their results using the dressed propagator approach for $j=2$, extended to $j=0$ in \cite{Calmet:2013hia}. Here we present the 1-loop corrected dressed graviton propagator as follows,
\begin{equation}\label{dress}
i \mathcal{D}^{\alpha \beta, \mu \nu}= \frac{i}{2 q^2}\left(1+2 B\left(q^2\right)\right)\left[L^{\alpha \mu} L^{\beta \nu}+L^{\alpha \nu} L^{\beta \mu}-L^{\alpha \beta} L^{\mu \nu}\right] -i \frac{A\left(q^2\right)}{4} L^{\alpha \beta} L^{\mu \nu} .
\end{equation}
where
\begin{align}
    A(q^2)=&-\frac{1}{30\pi}G_N N_s (1+10\lambda+30\lambda^2)\ln{\left(\frac{-q^2}{\mu^2}\right)},\\
    B(q^2)=&\frac{1}{240\pi}G_N N_s\ln{\left(\frac{-q^2}{\mu^2}\right)},\\
    L^{\mu \nu}=&\eta^{\mu\nu}-\frac{q^\mu q^\nu}{q^2},
\end{align} 
where $q$ represents momentum and $\mu$ is related to the renormalization scale of the theory. Base graviton propagator can be recovered by setting $A(q^2),\ B(q^2)=0$. Again, setting $\lambda=0,-\frac{1}{3}$ returns the $j=2$ dressed graviton propagator, and the $\lambda$ dependent terms correspond to the off-shell $j=0$ part. Authors in \cite{Aydemir:2012nz} mistakenly assume that $A(q^2)$ goes to zero completely in the aforementioned limits and due to this, their result for the infinite 1-loop summed dressed propagator is incorrect.

Also, in order to proceed with the summation, they assume that $G_N N_s$ is small. Since this is a dimensionful quantity, a more complete statement would be $G_N N_s\ll s^{-1}$. This again gives us a similar upper bound on energy as (\ref{bound}), (\ref{unlambound}), meaning its dependence on $N_s$ is still present. Further, they only proceed with the $j=2$ part for the rest of the calculation, which as stated earlier, doesn't hold any information about the violation scales dependent on $\lambda$.

The authors of \cite{Calmet:2013hia} come to the rescue here. They assume $\lambda_1=\lambda_2=\lambda$ (i.e. small Higgs' background limit) and focus on the $j=0$ part of the dressed propagator (\ref{dress}) by assuming $\lambda\gg1$, which is a valid limit as mentioned earlier in this section. This limit is suggested by the authors to avoid any contributions from $B(q^2)$, and consequently also ignore parts of the base propagator contribution. The dressed propagator in this work looks like,
\begin{equation}
    i \mathcal{D}^{\alpha \beta \mu \nu}\approx-\frac{i}{2 q^2}\left(1+\frac{q^2A(q^2)}{2}\right) L^{\alpha \beta} L^{\mu \nu}\approx-\frac{i}{2 q^2}\left[1-\frac{q^2 G_N N_s\lambda^2}{2\pi}\ln{\left(\frac{-q^2}{\mu^2}\right)}\right] L^{\alpha \beta} L^{\mu \nu}.
\end{equation}
Typographical errors aside, these are the results of \cite{Calmet:2013hia}. Later, similar to \cite{Aydemir:2012nz}, they assume $\lambda^2 G_N N_s\ll q^{-2}$ to be able to sum over the infinite series (which again reinforces the dependence of the perturbative unitarity bound on $N_s$ and $\lambda$ as in (\ref{lambound}) and (\ref{unlambound1})). This, however, doesn't make sense because even though taking the two aforementioned limits simultaneously means that we can ignore $B(q^2)$ (since $\lambda\gg1$) in favour of $A(q^2)$ in (\ref{dress}), all parts of the base graviton propagator must contribute to the dressed result since the constraint $q^2G_N N_s\lambda^2\ll1$ implies that $A(q^2)$ is the leading perturbative correction. The actual form of the propagator, as per their assumptions, should look like,
\begin{equation}\label{1ldressed}
i \mathcal{D}^{\alpha \beta, \mu \nu}\approx \frac{i}{2 q^2}\left[L^{\alpha \mu} L^{\beta \nu}+L^{\alpha \nu} L^{\beta \mu}-L^{\alpha \beta} L^{\mu \nu}\right] -i \frac{G_N N_s\lambda^2}{4\pi}\ln{\left(\frac{-q^2}{\mu^2}\right)} L^{\alpha \beta} L^{\mu \nu},
\end{equation}
for which summing the infinite series is a rather difficult task. Therefore, we need to confine ourselves with the 1-loop result in this method as well. It can be verified that the partial wave amplitude for the sum of $\phi_A\phi_A\to\phi_B\phi_B$ type processes involving the 1-loop dressed graviton propagator (\ref{1ldressed}) for all $A,\ B$ is the same as that obtained in (\ref{1lns}) assuming $\lambda\gg1$ (up to a multiplicative factor), i.e. $|a_0^{tree}|^2=\text{Im}(a_0^{1-loop})$ as verified using the partial wave amplitude approach in \cite{Calmet:2013hia}.

\section{Discussion}\label{concl}


The self-healing mechanism was defined in \cite{Aydemir:2012nz} to operate under specific conditions that were first discovered in \cite{Han:2004wt} and listed in Sec.\ref{prop}. After examining the claims made in the paper in both partial wave and dressed propagator approaches, we conclude through this work that the assessment made by the authors of \cite{Han:2004wt} that the unitarity violation scale depends on the number of particles $N_s$ is indeed true and complete healing of tree-level violations works only if the bounds described in (\ref{unlambound}) and (\ref{lambound}) are obeyed strictly.

In conclusion, we found that tree-level unitarity violations are indeed healed using 1-loop corrections, but the conditions required to effectively apply the self-healing mechanism themselves impose bounds on the energy scale of the theory that are dependent on the number of particles in the theory, as found by the authors of \cite{Han:2004wt} previously.

\section*{Appendix}\label{appendix}
The authors in \cite{Aydemir:2012nz} claim that for a unitary theory obeying $\text{Im}(a^{leading}_j)=|a^{tree}_j|^2$, we can write,
\begin{equation}
    a_j=a_j^{tree}+a_j^{1-loop}=\frac{a_j^{tree}}{1-\frac{\text{Re}(a_j^{1-loop})}{a_j^{tree}}-ia_j^{tree}}.
\end{equation}
Expanding $a_j^{1-loop}=\text{Re}(a_j^{1-loop})+i\text{Im}(a_j^{1-loop})$, we find that the equation above holds true if and only if,
\begin{equation}
    \text{Re}(a_j^{1-loop})^2+2a_j^{tree}\text{Im}(a_j^{1-loop})+(a_j^{tree})^4=0
\end{equation}
Real solutions exist for $\text{Re}(a_j^{1-loop})$ only when $|a_j^{tree}|\leq1$. 

\bibliographystyle{unsrtnat}
\bibliography{refs}

\end{document}